\documentclass[journal=jacsat,manuscript=article]{achemso}

\usepackage[version=3]{mhchem} 
\usepackage{xcolor}

\author{C. Granados}
\affiliation{Department of Physics, Guangdong Technion - Israel Institute of Technology, 241 Daxue Road, Shantou, Guangdong, China, 515063}
\email{camilo.granados@gtiit.edu.cn}
\alsoaffiliation{Technion - Israel Institute of Technology, Haifa, 32000, Israel}
\alsoaffiliation{Guangdong Provincial Key Laboratory of Materials and Technologies for Energy Conversion, Guangdong Technion - Israel Institute of Technology, 241 Daxue Road, Shantou, Guangdong, China, 515063}

\author{B. Kumar Das}
\affiliation{Department of Physics, Guangdong Technion - Israel Institute of Technology, 241 Daxue Road, Shantou, Guangdong, China, 515063}
\alsoaffiliation{Technion - Israel Institute of Technology, Haifa, 32000, Israel}
\alsoaffiliation{Guangdong Provincial Key Laboratory of Materials and Technologies for Energy Conversion, Guangdong Technion - Israel Institute of Technology, 241 Daxue Road, Shantou, Guangdong, China, 515063}

\author{Christian Heide}
\affiliation{Stanford PULSE Institute, SLAC National Accelerator Laboratory, Menlo Park, California 94025, USA}
\alsoaffiliation{Applied Physics, Stanford University, Stanford, CA, 94305, USA}

\author{Shambhu Ghimire}
\affiliation{Stanford PULSE Institute, SLAC National Accelerator Laboratory, Menlo Park, California 94025, USA}

\author{M. F. Ciappina}
\email{marcelo.ciappina@gtiit.edu.cn}
\affiliation{Department of Physics, Guangdong Technion - Israel Institute of Technology, 241 Daxue Road, Shantou, Guangdong, China, 515063}
\alsoaffiliation{Technion - Israel Institute of Technology, Haifa, 32000, Israel}
\alsoaffiliation{Guangdong Provincial Key Laboratory of Materials and Technologies for Energy Conversion, Guangdong Technion - Israel Institute of Technology, 241 Daxue Road, Shantou, Guangdong, China, 515063}

\title[Attosecond vortices in semiconductor materials]
  {Attosecond vortices in semiconductor materials}

\begin{document}


\begin{abstract}
We present the first theoretical results on the generation of short-wavelength attosecond vortex beams in semiconductors through their interactions with an intense Laguerre-Gauss beam, in the limit where non-perturbative high-order harmonics are generated. We exploit the details of the novel microscopic mechanism for high-order harmonic generation (HHG) in condensed matter, such as the use of dephasing time included in semiconductor Bloch equations (SBE), the combination of the SBE model with the thin slab model, and the use of experimentally verified scaling laws for various harmonic orders. For our test, we use a zinc oxide crystal as our standard sample, and our vortex beam is characterized by a topological charge of $l=1$. Our time-domain analysis shows that harmonics within the plateau region specifically contribute to the generation of the attosecond vortex beam. Our findings have implications for advancing the understanding of solid-state HHG and leveraging its strengths, such as the use of thin and dense media, for the efficient generation of short-wavelength attosecond vortex beams.   

\end{abstract}

\section{Introduction}

The observation of high-order harmonics in semiconductor crystals subjected to intense mid-infrared laser fields \cite{Ghimire} and subsequent advances have launched several exciting possibilities at the intersection of materials science and attosecond photonics such as the manipulation of the electronic band structure of the bulk media, spectroscopic probing of the topologically protected surface sates, and the generation of attosecond pulses in the compact setups.  The use of solids as the source material comes with some advantages over their atomic gas counterparts such as their high efficiency as their HHG threshold peak intensities are modest \cite{sources}. Also, semiconductor materials, in combination with beams carrying orbital angular momentum, e.g.~vortex beams, could potentially be used to shape the outgoing short wavelength light \cite{Garcia2024}.  Furthermore, the use of mid-infrared laser fields as the driving fields bring the spectral window of plateau harmonics to wavelengths longer than 200 nm, which does not require the use of vacuum apparatus\cite{Chang2022}.

High-harmonic generation (HHG) in solid-state materials can be explained by the evolution of electrons and holes within the material's bands under the influence of a strong driving-laser field \cite{Vampa14,Vampa15}. Initially, an electron in the valence band is accelerated by the driving-laser field towards the $\Gamma$-point ($k=0$) between the valence and conduction bands. At the $\Gamma$-point, where the energy difference between the bands is minimal, the corresponding transition probability is maximized. With increasing amplitude of the driving-laser field, the electron is excited to the conduction band via either multi-photon absorption or tunneling, creating a hole in the valence band. Subsequently, the electron-hole pair evolves within their respective bands under the influence of the driving-laser field. As the amplitude of the driving-laser field decreases, the electron-hole pair returns to the vicinity of the $\Gamma$-point, where recombination occurs. Notably, unlike in atomic gases, the recombination of electrons and holes in solid materials may occur at a position different from the initial position of the electron's escape \cite{Vampa15}. The excess energy imparted to the solid material by the external field is released in the form of high harmonic radiation. Added to this, the bandgap of the semiconductor material (significantly smaller than the ionization potential of atoms ) limits the energy of the emitted harmonics. It is the interplay between the laser intensity and the materials bandgap that limits the emitted harmonics. The bandgap is limited by the used material and the laser intensity by the damage threshold. 

The quantitative description of the HHG process in solid materials is mainly realized by three different theoretical models: the time-dependent Schr\"odinger equation (TDSE) \cite{TDSE1,TDSE2,TDSE3}, the semiconductor Bloch equations (SBE) \cite{SBE1, SBE2, SBE3, Haug09}, and the time-dependent density functional theory (TDDFT) \cite{Runge84,TDDFT1,TDDFT2}. However, a unified theory elucidating the intricate mechanism for HHG in solids is still an open question. The difficulty in developing a general model for HHG in solids, contrary to the atomic case, may lie in the fact that solids present different symmetries \cite{Ghimire3}, correlations, and regimes of coupling with the laser field. Additionally, it has recently been observed that differences between the harmonics generated via reflection or transmission configurations can lead to the detection of a material-dependent dipole phase, which can be used for harmonic spectroscopy in solids \cite{Ghimire2}. These differences manifest in different selection rules for HHG, for example \cite{Oren23}. Adding to the complexity of the target material, many different laser fields can be used to retrieve information from the target \cite{Oren23} or shape the XUV field into novel structures via non-trivial light spin-orbit couplings \cite{Garcia2024}.

Despite the differences between HHG in solids and atomic gases, we recently demonstrated that, similar to gas-phase HHG, the generation of high harmonic vortices in semiconductor materials can be mathematically formulated using the thin slab model (TSM) in combination with the SBE \cite{Granados1}. The atomic counterpart describes the interaction of the vortex beam with a thin slab of atoms within the framework of the strong-field approximation (SFA) \cite{Lewenstein2}. For the solids case, the TSM restricts the generation of harmonics to the last layer of the solid materials, which has been experimentally demonstrated for atomically thin MoS$_2$ \cite{Ghimire3}. Working with monolayer materials presents interesting physical differences compared to bulk materials, such as different symmetries, enhanced many-body effects and the absence of propagation effects inside the material \cite{Ghimire3}. This scenario prompts inquiry into the effect of the spatio-temporal symmetries, of the vortex driving field, in the selection rules arising from the interaction with a semiconductor monolayer, as MoS$_2$. Additionally, materials like MoS$_2$ and graphene serve as promising candidates to investigate harmonic generation driven by vortex beams, given the routine preparation of monolayers of these materials in laboratories. Moreover, optical vortex beams for HHG are currently experimentally feasible \cite{Gauthier19}, further facilitating the research in this domain. This creates the perfect platform to test the validity of the TSM and its predictive power in solid materials. Importantly, the expected larger flux of harmonic light because of the larger atomic density in solids make it possible to characterize the higher order XUV-vortex beams, in contrast to the atomic case, where the low flux of XUV-vortex hampers its characterization \cite{SingBeams}. Similarly, the use of perfect optical vortex beams in solid state materials will allow for the full characterization of the XUV-vortices, since the divergence of the POV beams is independent of the topological charge.  

Optical fields carrying orbital angular momentum (OAM) \cite{Corkum, Geneaux, Kong2019, Gauthier:2017, CarlosPRL}, or vortex beams, offer a powerful means to manipulate light-matter interactions on a microscopic scale. These beams find applications in diverse fields such as super resolution microscopy \cite{Wang17}, photolithography \cite{Li2018}, and molecular chirality detection \cite{Forbes2019,Ward2016}. Another example of their applicability in the XUV and soft X-ray spectral regimes lies in the fact that vortex beams present unique advantages, including enhancing resolution in focusing and imaging systems, as well as enabling XUV interference lithography for nanostructure fabrication \cite{Becker2011}. In particular, nanostructured samples such as Fresnel zone plates can offer further opportunities for XUV focusing compared to the gas phase \cite{Korobenko2022}. Additionally, as in the atomic case where the law of OAM upscaling \cite{Paufler2019} is preserved and harmonic vortices are generated with similar divergence, recent experimental work \cite{Gauthier19} demonstrates the feasibility of driving HHG in solids with vortex beams, revealing a consistent OAM scaling and uniform harmonic divergence. We demonstrated that a multiscale model, combining the SBE with the TSM \cite{Granados1}, accurately reproduces these experimental findings. Moreover, HHG driven by vector beams from monolayer graphene \cite{Garcia2024} results in a unique necklace pattern formation in the far-field. However, to date, investigations into attosecond vortex generation in semiconductor materials remain limited. Realizing attosecond structured light capable of interacting with the target system angular momentum holds promise for understanding spin-orbit coupling effects in solids and their influence on HHG. Additionally, generation of perfect optical vortices in solid materials with higher angular momentum and a high flux of photons, opens the door to increase the resolution in stimulated emission depletion (STED) microscopy \cite{SuperResol}. 

In this contribution, we initially validate our theoretical multiscale approach for HHG in solids by employing a combination of the TSM and the SBE. Our validation is based on comparing the theoretical scaling law of the harmonics amplitude, as a function of the driving laser intensity, with experimental results.
Secondly, we delve into a theoretical exploration of the feasibility of generating attosecond vortex beams through HHG in ZnO. We achieve this by synthesizing multiple high harmonic orders.

\section{Theoretical description}

In this section, we will discuss the implementation of the thin slab model and the semiconductor Bloch equations to calculate the far-field response from the ZnO crystal. We will begin by introducing the spatio-temporal driving laser field, followed by the SBE model, and then present the calculations of the far-field complex amplitude for harmonic vortices.

\subsection{Spatio-temporal vortex beam}
The spatio-temporal complex field amplitude of a linearly polarized LG driving-laser beam can be written as  \cite{Granados1}: 

\begin{eqnarray}
\label{lgbeam}
    U(r',\phi',z,t) &=& U(r',\phi',z) E(t) \nonumber \\ &=& \Bigg[\frac{\omega_0}{\omega(z)}\Bigg(\frac{\sqrt2 r'}{\omega(z)}\Bigg)^{l} e^{-\Big(\frac{r'}{\omega(z)}\Big)^{2}} e^{i l \phi'} L_{P_0}^{l}\Bigg( \frac{2(r')^2}{\omega^2(z)} \Bigg) e^{i \kappa z}e^{\frac{i \kappa\; (r')^2}{2R(z)}}e^{i\varphi_{G}(z)}\Bigg]E(t),
    \label{STF}
    \end{eqnarray}

where $l$, $P_0$, $\omega_{0}$, $\varphi_G=-(2P_0+l+1)\arctan(z/z_R)$, and $z_R=\frac{1}{2}k\omega_{0}^{2}$ represent the orbital angular momentum, the radial index, the Gaussian beam waist, the Gouy phase, and the Rayleigh range of the LG beam, respectively. The phase front radius is defined by $R(z)=z\left[1+\left(z_R/z\right)^2\right]$, and the width of the beam at some propagation distance $z$ is $\omega(z)=\omega_{0}\sqrt{1+\left(z/z_R\right)^2}$. Now, the temporal part of the LG beam, $E(t)$, is defined as a laser pulse with a $\sin^2$ envelope:

\begin{eqnarray}
    \label{et}
    E(t)&=&E_0\text{sin}^2\Bigg(\frac{\pi t}{n_c T}\Bigg)\text{sin}(\omega_L t),
\end{eqnarray}
where $E_{0}$ represents the peak electric field amplitude, $n_c$ is the number of laser field cycles, and $T = 2\pi/\omega_L$ is the period of the driving-laser field, with $\omega_L$ being its corresponding frequency. 

Incorporating the spatio-temporal field described in Eq.~(\ref{STF}) into the SBE model is extremely complicated and should be avoided. To achieve this, we first consider that the contribution to the total harmonics measured at the detector comes from the harmonics generated in the last few layers of the solid target~\cite{Ghimire}. Second, we assume that the dipole approximation remains valid. This assumption is supported by the short excursion distance of the electron compared with the long wavelength of the driving-laser field ($\lambda=3.25$ $\mu$m). The maximum electron excursion values can be computed as $r_{\text{max}}=eE_0\lambda^2/4\pi^2mc^2 = 32$ \textup{~\AA} \cite{Ghimire}. In our case, the maximum considered driving-laser field amplitude results in an excursion distance $r_{\text{max}}=6.4$ \textup{~\AA}. In addition, we have considered in Eq.~(\ref{STF}) a zero radial index ($P_0 = 0$), which led to $L^l_{P0}(...) = 1$. This ensures an input LG beam profile with only one radial ring, thereby removing the intensity radial dependency from more transverse rings in the solid-HHG calculations.

By using these two approximations, we can calculate the harmonic spectra by considering only the temporal part of $U(r',\phi',z,t)$, i.e., $E(t)$ in the SBE model, as will be shown in the following subsection.

\subsection{Semiconductor Bloch equations}

Solid-state high harmonic generation (Solid-HHG), particularly the interband contribution to the process, follows a conceptual framework resembling to the three-step model observed describing HHG in atomic systems \cite{Vampa14}. While this schematic elucidates the generation of harmonics above the material's band gap, it's worth noting that below the bandgap the intraband mechanism can also play an important role. In the generation of harmonics below the band gap electrons oscillate within the same energy band at high frequencies, leading to the emission of harmonic radiation.
In the interband contribution, the electron-hole dynamics are primarily governed by the microscopic polarization (coherence), denoted as \( p_k \), between the valence and conduction bands. The intraband contribution is dominated by variations in electron (or hole) occupation (intraband current, $j(t)$ within the same band. The theoretical description of these two mechanisms leading to solid-HHG is encapsulated in the Semiconductor Bloch equations (SBE). The SBE describing the electron-hole dynamics have the following form:
\begin{eqnarray}
\label{sbes}
\nonumber
i\frac{\partial}{\partial t}p_k&=&\left(\varepsilon_k^e+\varepsilon_k^h-i\frac{1}{T_2}\right) p_k-(1-n_k^e-n_k^h)E(t)d_k\\
&&-iE(t)\nabla_kp_k, \nonumber\\
\frac{\partial}{\partial t}n_k^e&=&-2\mathrm{Im}[ E(t)d_k p_k^{*}]-E(t)\nabla_k n_k^e,\\
\frac{\partial}{\partial t}n_k^h&=&-2\mathrm{Im}[ E(t)d_k p_k^{*}]-E(t)\nabla_k n_k^h \nonumber,
\end{eqnarray}
where $\varepsilon_k^{e(h)}$ are the single particle energies of the electrons (holes), $n_k^{e(h)}$ is the occupation of electrons (holes), $T_2$ is the dephasing (or, decoherence) time of the polarization. The dipole transition matrix element ($k$-dependent) between the valence and conduction band is represented by $d_k$. For the numerical solution of Eqs.~(\ref{sbes}), we assume a real dipole transition matrix with a constant value of $d_k$ at the $\Gamma$ point~\cite{PhaseSolids}. 

From Eqs.~(\ref{sbes}), we can compute the total time-dependent interband polarization $P(t)$ and intraband current density $J(t)$~\cite{Trung16}. The combination of the two observables gives rise to the total emitted spectral intensity, $S(\omega)$:

\begin{eqnarray}
S(\omega)\propto|\omega\;P(\omega)+iJ(\omega)|^2.
\end{eqnarray}

Since we are interested in testing our theoretical approach at the nonlinear region, we will consider only the contribution of the interband polarization $P(\omega)$. For this region the intraband contribution to the solid-HHG is negligible \cite{SolidTutorial}. We will also consider here a driving-laser field with linear polarization along the $\Gamma-K$ crystal direction and a one-dimensional crystal. These restricts the SBE to a one dimensional model. Even when our model is one dimensional, our approach allows for testing the generation of harmonics for different crystal directions, just by changing the orientation of the material with the respect to the laser field polarization. 

The SBE were solved using a driving-laser field wavelength of $3.25$ $\mu$m and for a range of peak laser intensities ($I_0 = 1-1.5$ TW/cm$^2$) interacting with a ZnO crystal oriented along the $\Gamma$-K direction. The number of optical cycles was set to $n_c = 9$, corresponding to a total pulse duration of approximately 97 fs. The value of the dephasing time was set to $T_2 = 1.0$ fs. The choice of dephasing time will be justified in the results sections when the theoretical scaling laws are compared with the experimental results. From the set of calculated spectra, we can extract the amplitude scaling law of the different harmonics as a function of the fundamental laser amplitude. In our model, the bandgap energy of ZnO is set to $E_g = 0.1213$ a.u. (3.3 eV). The transition dipole moment matrix is assumed to be real and constant, with a value of $d_k = 3.46$ a.u. (corresponding to value at the $\Gamma$-point~\cite{PhaseSolids}). For solving the SBEs, we use two bands: one valence band and one conduction band, defined by: 
\begin{eqnarray}
    E_c(k) &=& \sum_j \alpha_{v}^j\cos({j k a})\nonumber \\
    E_v(k) &=& E_g + \sum_j \alpha_{c}^j\cos({j k a})\nonumber\\
    \varepsilon_g(k) &=& E_c(k)-E_v(k). 
\end{eqnarray}
Here, $\varepsilon_g$ represents the bandgap energy for each value of momentum $k$. The coefficients of the expansion, $\alpha_{v,c}$, are obtained by expanding the bands within the tight-binding approximation \cite{Haug09}. The lattice constant $a = 4.61$ is defined for the $\Gamma-K$ direction.

The solution of the SBE equation for different values of fundamental laser field, allows us to extract the scaling factor $p$ for each harmonic. This parameter is fundamental to scale the microscopic response of the solid target to the far field, as will be shown in the following sections. 


\subsection{Harmonic vortices at the far-field}

Armed with the amplitude scaling law ($p$-value), we solve the Fraunhofer diffraction integral for each individual harmonic case using the TSM. This allowed us to calculate the expression for the far-field complex field amplitude, $A_q^{f}(\beta, \phi)$, profile as a function of the harmonic order, resulting in:

\begin{eqnarray}
    A_q^{f}(\beta, \phi)&\propto& i^{ql} e^{iql\phi} \int_0^\infty r' dr' A^{n}_{q} J_{ql}\Bigg( \frac{2\pi\beta r'}{\lambda_q} \Bigg),  \label{far}
\end{eqnarray}

with the near-field complex amplitude given by $A^{n}_{q} \propto e^{iql\phi^{'}}\big|U(r')\big|^p$ , and $p$ represents the scaling factor. Here, $\beta$ and $\phi$ represent the divergence and the azimuthal coordinate in the far-field, respectively. $\lambda_{q}=\lambda/q$ denotes the wavelength of the $q^{\mathrm{th}}$ harmonic order, and $J_{ql}(...)$ represents the Bessel function of order $ql$. Generally, the amplitude of the $q^{\mathrm{th}}$ harmonic order, $U_q$, scales with the fundamental field amplitude, $U_f$, depending on the interaction regime with the driving-laser. In the perturbative regime, the scaling law, $U_q\propto U_{f}^p$, is satisfied with $p=q$. However, in the non-perturbative regime the relation between the harmonic order and the p-values is $p<q$. Similar to the atomic HHG case \cite{BikashInProgress}, the OAM of the $q^{th}$ harmonic order scales as $l_q = q l$, which follows from the OAM conservation. This is not surprising since the only difference between the atomic and solid-state TSM is the harmonics amplitude scaling law. By positioning the target at $z=0$, i.e., at the laser focus, and disregarding the dipole phase, Eq.~(\ref{far}) can be solved analytically to calculate the intensity of harmonic vortices in the far-field, given by:

\begin{eqnarray}
    I(\beta_x,\beta_y)&\propto&\frac{I_0^p (1)^{ql}\left(\frac{2 \pi q \sqrt{\beta_x^2+\beta_y^2}}{    {\lambda_0}}\right)^{2ql}\Gamma \left[\frac{1}{2} (pl+ql+2)\right]^2}{2^{2ql+2}\left(\frac{p}{\omega_0^2}\right)^{l p+l q+2}\Gamma[ql+1]^2}\nonumber\\
    &\times& \left\{_1F_1 \Bigg[\frac{pl+ql+2}{2};ql+1;-\frac{\pi^2 q^2 \omega_0^2 \left(\beta_x^2+\beta_y^2\right)}{p\lambda_0^2} \Bigg]\right\}^2.\label{TIP}
\end{eqnarray}

\section{Results and discussion}

In this section, we start with a concise overview of the experimental setup, followed by a comparison between the experimental and theoretical findings. This comparison serves as a validation test for our multiscale model. With the robustness of our theoretical approach established, we then delve into exploring the potential for generating attosecond twisted pulses. This entails synthesizing a segment of the calculated harmonic spectra corresponding to the measured harmonic orders.

The experimental procedure for measuring the harmonics generated in ZnO has been previously published, and we refer the reader to Ref.~\cite{Ghimire} for more details. Here, we use the experimental results to benchmark our approach. In summary, a Gaussian-shaped driving-laser beam with wavelength centered at 3.25 $\mu$m, a pulse duration of approximately 100 fs (around 9 cycles) was employed to generate harmonics in a 500 $\mu$m thick crystal cut perpendicular to its optical axis. The maximal applied intensity at the sample was about 5 TWcm$^{-2}$. The generated harmonics were focused onto a spectrometer capable of measuring wavelengths down to 155~nm. The amplitude scaling of the high harmonics (9$^{\text{th}}$ - 21$^{\text{st}}$) as a function of the fundamental driving-laser field amplitude in the range of approximately 0.28 to 0.6 V/\AA\, is plotted in Fig.~\ref{fig3}(a). While there are demonstrations of harmonic (3rd to 9th order) beams carrying OAM generated using LG beams~\cite{Gauthier19}, to the best of our knowledge, there is no experimental demonstration of high-order harmonics (above 11th order) beams carrying OAM generated using solid samples. These high-order harmonics are a fundamental requisite to generate attosecond pulses~\cite{ParisInProgress}.


\subsection{Harmonics amplitude scaling law}

The knowledge of the harmonic amplitude scaling law is a prerequisite for constructing the harmonic vortices spatial profile in the far-field~\cite{Granados1}. Experimentally, this scaling can be measured by changing the driving-laser field amplitude and recording the changes in a set of harmonics amplitude, as described previously. Theoretically, however, several parameters can change the general trend of the harmonics amplitude scaling law \cite{SBEintensity, IntenDependSBE}. One of them is the dephasing time, $T_2$. Here, we set its value to $T_2 = 1.0$ fs to reproduce the linearity between the harmonics amplitude and the driven laser field peak field strength, observed in the experiment~\cite{Ghimire}. Notice that previously, the SBE model predicted oscillation dependent scaling laws \cite{IntenDependSBE}, however, here we prove that the experimental dependency is linear and can only be reproduce theoretically with the proper dephasing time. Higher dephasing time values resulted in amplitude scaling laws for individual harmonics which differ significantly from each other and shown oscillatory behaviour \cite{IntenDependSBE}. This in turn can not reproduce the experimental linear trend which leads to similar amplitude scaling law values. We shown in the supplementary material to this work, the scaling laws for the different de-coherence times as well as for different crystal directions. We also need to highlight that recently it has been demonstrated theoretically, the need for using extremely short dephasing times in semiconductors \cite{DephaSemi}. 

\begin{figure}[h!]
\includegraphics[width=0.9\textwidth]{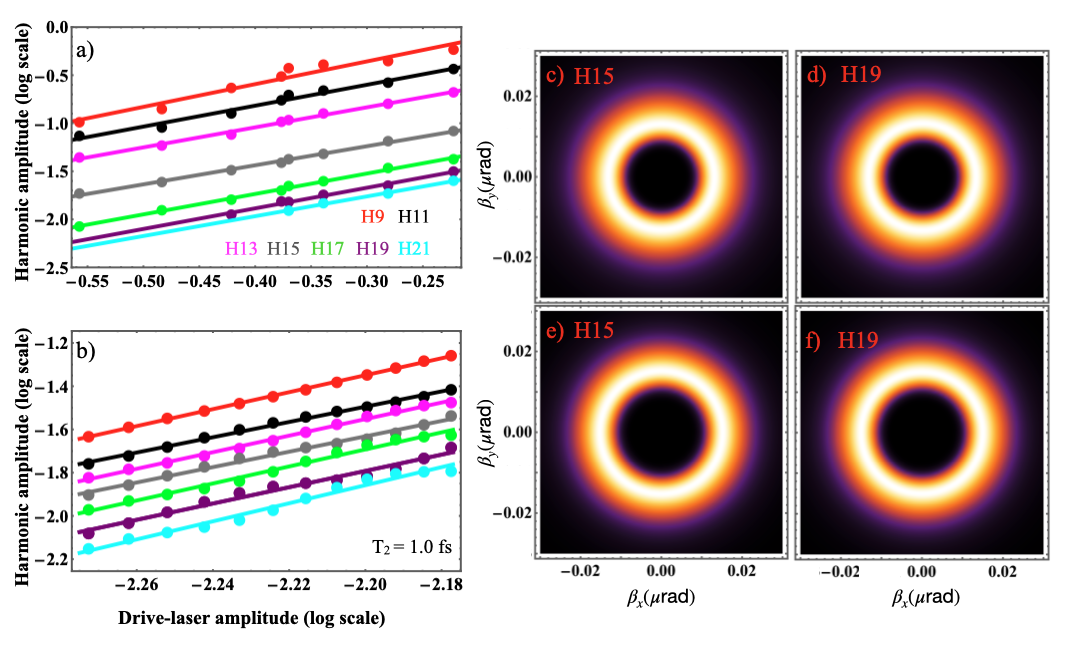}
\caption{Experimental and theoretical amplitude scaling laws for different harmonic orders. (a) experimental scaling laws measured with a Gaussian intensity profile. In (b) theoretical scaling laws calculated with a $\text{sin}^2$ intensity profile. In (c) and (d) harmonic vortex spatial profiles calculated with the experimental scaling laws for harmonics 15$^{\text{th}}$ and 19$^{\text{th}}$, respectively. In (e) and (f) same as in (c) and (d), but using the theoretical scaling laws and the $\Gamma-K$ crystal direction. 
}\label{fig1} 
\end{figure} 

In Fig.~\ref{fig1} we show the comparison between the scaling laws for the experimental measurements in (a) and theoretical calculations in (b) using Eq.~(\ref{sbes}). The dots in both panels (a) and (b), represent the experimentally measured and the theoretical values of the harmonics amplitude as a function of the driving-laser field peak amplitude. The calculations were performed with a intensity profile described by $\text{sin}^2$ time envelope, as shown in Eq.~\ref{et} and the $\Gamma-K$ crystal direction. Calculations for to the $\Gamma-M$ direction show similar results and can be found in the Supplemental Material. Overall, both calculations show the necessity of small decoherence time to reproduce the experimental amplitude scaling law. The corresponding lines represent the linear fit to the data used to extract the amplitude scaling law values. The amplitude scaling laws extracted from the experimental data for harmonic order ranging from 9$^{\text{th}}$ to 21$^{\text{st}}$ are \{$2.4, 2.2, 2.1, 2.0, 2.1, 2.1, 2.1$\}. Likewise, the theoretical values were found to be \{$ 3.9, 3.5, 3.8, 3.6, 3.9, 3.8, 4.2 $\}.  Notice that, even when the experimental and theoretical amplitude scaling laws values are different, the latter are about a factor 2 of the former, the general trend is well-reproduced by the SBE model.

In Figs.~\ref{fig1}(c) and (d), we present the calculated spatial profiles of the harmonic vortices of order 15$^{\text{th}}$ and 19$^{\text{th}}$, respectively. These profiles were calculated with the amplitude scaling laws extracted from Fig.~\ref{fig1}(a). Furthermore, in (e) and (f), we present the spatial profiles for the same harmonic orders as in c) and d), but calculated with the amplitude scaling laws extracted from Fig.~\ref{fig1}(b). Here, for the spatial profiles presented in Fig.~\ref{fig1} c) and d), the amplitude scaling law corresponds to $p_{H15} =  2.0$,  $p_{H19} = 2.1$. On the other hand, the theoretical amplitude scaling laws are $p_{H15} = 3.6$ and $p_{H19} = 3.8$. The differences in the amplitude scaling law for different harmonics, is reflected in slightly different vortices core size, which is larger for the theoretical calculations, as can be observed comparing Figs.~\ref{fig1}(c) to (d) with Figs.~\ref{fig1}(e) to (f).

\begin{figure}[h!]
\includegraphics[width=0.8\textwidth]{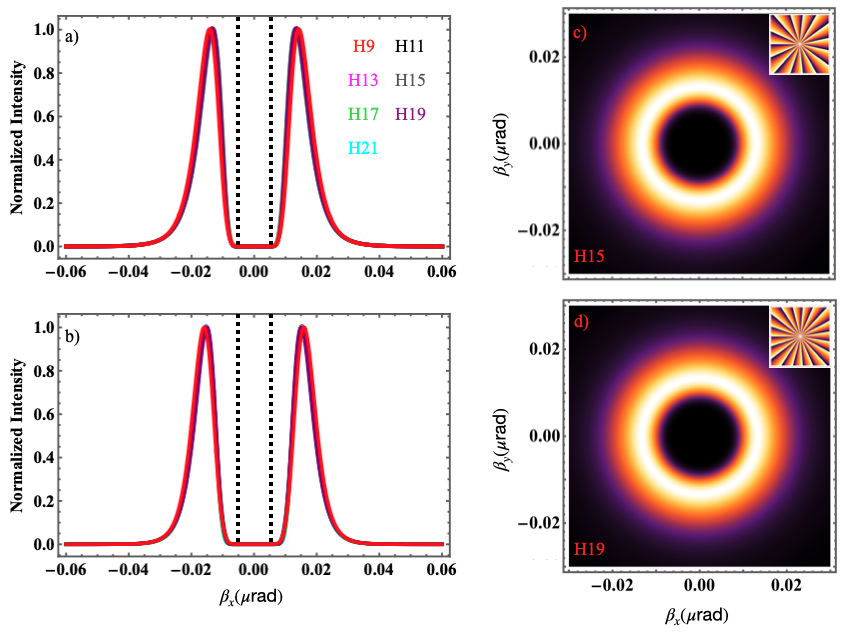}
\caption{Spatial line intensity profile and phase for the generated harmonic vortex. (a) line intensity profile for the plane $\beta_y = 0$ using the experimental scaling law.  (b) same as (a) but using the theoretical scaling law. Notice the small divergence of the different harmonic vortices. (c) and (d) complete spatial intensity profile for harmonics 15$^{\text{th}}$ and 19$^{\text{th}}$, respectively, using the experimental scaling law. In both cases, the inset plot depicts the harmonic vortex phase, which scales with the harmonic order, $q$. This is a direct consequence of the conservation of OAM for the generated harmonics. The black dashed lines serves only as a guide for the reader to compare the vortex core sizes.}\label{fig2} 
\end{figure}

In Fig.~\ref{fig2}, we present a more detailed characterization on the size change for different harmonic vortices. In Fig.~\ref{fig2} a), we present the line intensity profiles for different harmonics calculated for $\beta_y = 0$, and using the p-values extracted from Fig.~\ref{fig1} a) (experimental values). Likewise, in Fig.~\ref{fig2} b), we depict the line intensity profiles for $\beta_y = 0$, but this time with the p-values extracted from Fig.~\ref{fig1} b) (theoretical values). A negligible difference in the size of the generated harmonic vortices can be observed from both panels. This can be explained by the small changes between the scaling laws for the different harmonics (calculated either with theoretically or experimental scaling laws). Also notice, as presented in Figs.~\ref{fig1} c) to f), that the vortex core is different for the experimental and theoretical scaling laws. The experimental vortex core size is smaller than the theoretical one, explained by the fact that the experimental amplitude scaling law values are smaller than the theoretical ones. In Fig.~\ref{fig2} c) and d), we show the spatial profiles of the harmonic orders 15$^{\text{th}}$ and 19$^{\text{th}}$, respectively, extracted from the experimental scaling laws. Each panel is accompanied by the harmonic vortex phase. The harmonic vortex phase scales with the harmonic number which is a consequence of the conservation of OAM for the different harmonics i.e., $l_q = q l$, with $q$ the harmonic order. This can be seen in the far-field profile of the vortex beam described in Eq.~(\ref{far}), since the resulting harmonic vortex phase is given by $e^{iql\phi}$. 

It is important to highlight that in the non-perturbative regime studied here, the amplitude scaling laws for different harmonics are similar. This contrast with the perturbative regime \cite{Gauthier19,Granados1}, where the p-values for different harmonics scatter significantly. This was explained by the presence of different harmonic generation mechanisms \cite{Vampa15}, i.e, intraband for harmonic energies below the material's band gap, and interband for harmonic energies above the energy band gap \cite{Granados1}. Within the framework of the SBE model, a dephasing time of 2~fs was used to correctly describe the experimental behaviour of the low harmonic vortex \cite{Gauthier19}. However, it was found that the role of dephasing time in the SBE model is mostly irrelevant in the generation of low-order vortex harmonics. For the harmonics in the plateau region of the HHG spectra, a dephasing time of $T_{2} = 1$ fs is necessary to describe the general trend observed experimentally. Typically, in the SBE model, the dephasing time is responsible for electron-electron correlation and other multi-body effects and it is a complicated parameter to extract. By fixing a dephasing time value we are fixing the correlation matrix in the SBE model to be diagonal and with constant values. Since it depends directly on the microscopic polarization of the system, the dephasing time maps the strength and the length of correlation effects of the particular system under investigation and it has been shown that it does not depend on the frequency of the driving-laser field \cite{AndersonModel}. The dephasing time, however, appears to be sensitive to the driving-laser field amplitude changes, as we show here. In our approach, by fixing the dephasing time to reproduce the relative scattering of the experimental amplitude scaling laws, we demonstrate that, for the plateau region of the harmonic spectrum, the multi-body effects are different to the ones observed in the perturbative part of the spectrum (that seems to be insensitive to the dephasing time). A small dephasing time means that the contribution to the dynamics of the electron in the driving-laser field for the plateau region is dominated by the short electron trajectories \cite{Vampa15}. It is then pivotal to find phenomenological approaches that can limit the value of the dephasing time for different materials and for different laser matter interaction strengths. The vortex beam divergence is sensitive to the changes in the dephasing time, as demonstrated here, providing a simple yet elegant way to limit its value.

\subsection{Harmonics phase analysis}

The exact value of the dipole phase generated during the electron excursion between the valence and conduction bands in solid targets remains an open question from both experimental and theoretical perspectives. Two key studies, one experimental (Ref.\cite{Ghimire2}) and one theoretical (Ref.\cite{PhaseSolids}), have established two fundamental characteristics of electron dynamics and harmonic propagation in solids: (1) The transition dipole matrix can be assumed to be real for 1-D problems and (2) The phase accumulated by harmonics during high-harmonic generation in solids depends on whether the harmonics are generated in transmission or reflection geometries. In the reflection geometry, where harmonics are generated within the first few layers of the crystal, the process is dominated by the dipole phase \cite{Ghimire}. However, when harmonics are generated from the bulk of the crystal, propagation effects usually dominate \cite{Ghimire}. These propagation effects can be attributed to self-phase modulation (SPM), which occurs only in the fundamental beam and dominates the phase of the harmonics.

We can represent all the phases involved in the harmonic generation and propagation process for the far-field harmonic vortex as:
\begin{eqnarray}
    \exp(i\Delta \Phi) = \exp\large(iql\phi+i\alpha_j^S U_P/(\hbar \omega)+iqk r'^2/R(z)\large).
\end{eqnarray}
Here, the terms $\exp(iql\phi)$ and $\exp(i\alpha_j^S U_P/(\hbar \omega))$ correspond to the helical phase and the dipole phase of the $q^{th}$-order harmonic, respectively. The term $\exp\left(iqk r'^2/R(z)\right)$ represents the focal phase, which arises from the curved wavefronts of the beam. Since the harmonics are generated in the last layer of the solid in a transmission geometry, we can safely neglect the dipole phase \cite{Ghimire}, as will be shown below. For high-order harmonic generation in solids, the dipole phase was reported in Ref. \cite{Ghimire} to be: 
\begin{eqnarray}
     \phi_j=\alpha_j^S \frac{U_P}{\hbar \omega} = \alpha_j^S \frac{e^2}{2\hbar m_e c n \epsilon_0} \frac{I(z)}{\omega^3}.
\end{eqnarray}
Typical values of $\alpha_j^S$, a material-dependent coefficient, are around 1 for MgO and 0.4 for SiO$_2$ crystals in units of $\pi$~\cite{Ghimire}. This corresponds to a dipole phase of $0.15$ mrad for MgO and $0.057$ mrad for SiO$_2$. For both calculations, we assume a laser power of approximately $1$ TW/cm$^2$ and that the harmonics are generated in transmission geometry. As reported in Ref.~\cite{Ghimire}, the phase arising from propagation effects was calculated to be $0.6$ rad.

Assuming similar values of $\alpha_j^S$ for ZnO, the calculated dipole phase ranges from $0.86$ mrad to $2.3$ mrad. These values are still smaller compared to the calculated SPM phase, which is approximately 1.9 rad, as shown below.

The phase, $\Delta \phi_{NL}(\theta)$, due to SPM affects the harmonics indirectly: since the pump beam experiences spectral broadening due to SPM, the fundamental laser field becomes shorter in the time domain. This effect dominates the phase of all harmonics \cite{Ghimire}. The nonlinear phase accumulated by the fundamental field due to SPM can be represented as follows \cite{Ghimire}:
\begin{equation}
    \Delta \phi_{NL}(\theta) = \frac{2\pi}{\lambda}\int_0^L n_2(\theta)I(z)dz = \frac{2\pi}{\lambda} n_2(\theta)I_0 z_R\arctan\Bigg(\frac{L}{z_R}\Bigg).
\end{equation}
The nonlinear index of refraction for the ZnO crystal is $n_2(\theta) = 23\times10^{-13}$ esu \cite{NLC}. This value allowed us to calculate the nonlinear phase gained by the fundamental beam due to self-phase modulation (SPM) over a propagation distance of $L = 250$ $\mu$m, which was found to be approximately 1.9 rad, corresponding to a time delay of around 1.4 fs. Compared with reported values for MgO and SiO$_2$, this suggests that propagation effects do not significantly alter the duration of the fundamental beam. Consequently, only minimal changes in the duration of the harmonics are expected, indicating that the formation of attosecond pulses is feasible when electron dynamics are driven by ultrashort pulses in the ZnO crystal. 

In the next section, we will demonstrate the formation of twisted, or helical, attosecond pulses in the ZnO crystal based on the conclusions presented above. The generation of these helical attosecond pulses is consistent with the recent experimental demonstration of attosecond pulses in solids \cite{ParisInProgress}.   




\section{Attosecond twisted pulses}

The experimental confirmation of attosecond pulse generation in semiconductor materials represents a significant milestone in a longstanding pursuit \cite{ParisInProgress}. The theoretical framework provided by the SBE model not only validates this achievement but also opens avenues for exploring attosecond pulses theoretically in various contexts. In this section, our goal is to demonstrate the generation of attosecond vortex beams in semiconductor materials. Generating attosecond pulses in solids, akin to the atomic case, involves synthesizing multiple harmonic orders \cite{ParisInProgress, Lewenstein1}. While a driving-laser field with a homogeneous spatial distribution results in harmonics with similar spatial shapes, this characteristic typically plays a negligible role in attosecond pulse generation. Only the temporal aspect of the resulting harmonics is synthesized. However, to create a spatial vortex with an attosecond time duration, it is necessary to synthesize both spatial and temporal components. This entails ensuring that the spatial divergence of the harmonic vortices is similar and that there exists a long plateau in the harmonic spectra providing the necessary bandwidth for attosecond generation.

\begin{figure}[h!]
\includegraphics[width=1\textwidth]{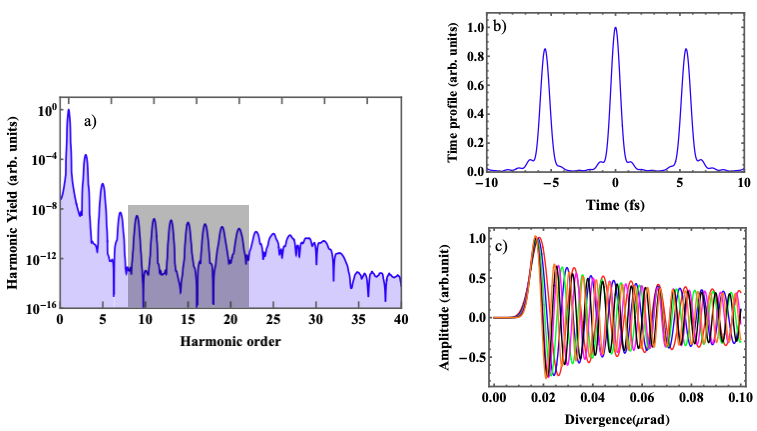}
\caption{Temporal and spatial characteristics of the generated high-order harmonics. (a) calculated harmonic spectrum in the ZnO crystal. The shadowed area depicts the harmonics used to generate the attosecond pulses. (b) Synthesized attosecond pulses. The central pulse time duration of the train, measured at the FWHM, results in 840 as. (c) the divergence of the different high-order harmonics. The low divergence around the maximum value of amplitude makes it possible to spatially synthesize the different harmonics. }\label{fig3} 
\end{figure}

In Fig.~\ref{fig3}, we illustrate the various spatial and temporal characteristics needed for generating an attosecond twisted pulse. 
In Fig.~\ref{fig3}(a), we present the harmonic spectrum calculated with the SBE for a laser intensity of 1.0~TW/cm$^2$. The shaded region within the harmonic spectrum represents the harmonics synthesized to generate the attosecond pulses. Here, we coherently combine the harmonics from the $9^{\text{th}}$ to the $21^{\text{st}}$, resulting in an attosecond pulse train with a central pulse duration of around $840$ as, as shown in In Fig.~\ref{fig3}(b). Since the duration of the attosecond pulse depends on the number of synthesized harmonics, theoretically, it is feasible to demonstrate the generation of isolated attosecond pulses of approximately 580~as by incorporating harmonics from the $9^{\text{th}}$ to the $31^{\text{st}}$ (not depicted here). 
Our calculations are based on the well-established case of atomic gases \cite{Lewenstein1}. The spatial divergence of the harmonic vortices is illustrated in Fig.~\ref{fig3}(c). It can be observed that the maximum amplitude of all harmonic orders, i.e., from the $9^{\text{th}}$ to the $21^{\text{st}}$, corresponds to nearly identical values of divergence, $\beta$. This, in turn, indicates that all vortex harmonics are emitted with almost identical divergence when HHG from solids is driven by optical beams carrying OAM. This results from the fact that the law of OAM upscaling is strictly obeyed during this nonlinear conversion process.

\begin{figure}[h!]
\includegraphics[width=0.8\textwidth]{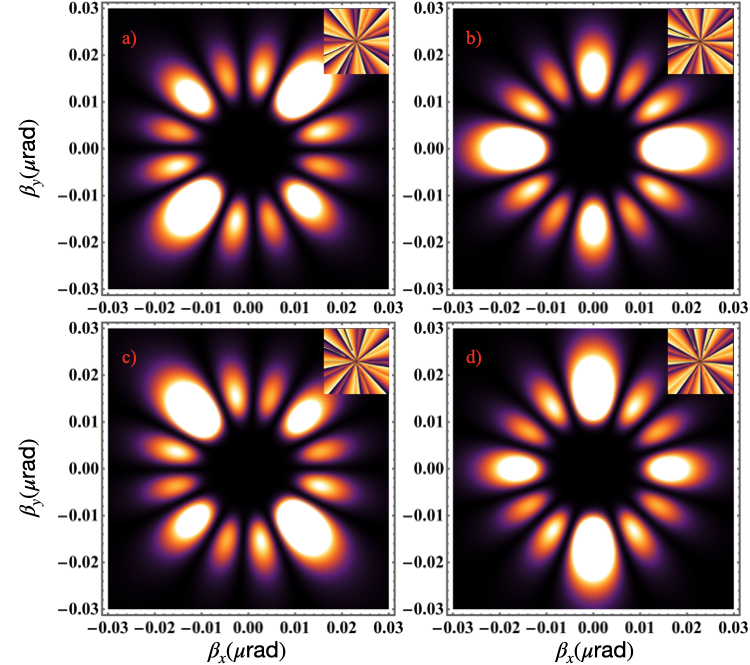}
\caption{Temporal evolution of the attosecond pulse spatial intensity distribution for different times within half a cycle of the laser period. (a) $\delta T=0.12 T$, (b) $0.25 T$, (c) $0.38 T$ and (d) $0.5 T$. The insets of all the panels corresponds to the phase of the harmonic vortex for the respective times.}\label{fig4} 
\end{figure}

The calculated spatial and temporal characteristics of the harmonic vortices showcase the potential for generating attosecond vortices. Another instrumental aspect of the vortex is its twisting nature. Thus, we analyze both the temporal evolution and the spatial distribution of the attosecond vortex beam. To do so, we incorporate the spatial and temporal phase terms from Eq.~(\ref{far}) and coherently combine the far-field amplitudes of individual harmonic vortices. This process yields a far-field amplitude for the attosecond twisted pulse that evolves over time:
\begin{equation}
    A_{as}^{f} = \sum_i^{Nh}  \Big|A_{q_i}^{f}(\beta, \phi)\Big|e^{iq_il\text{tan}^{-1}(\beta_y/\beta_y)}e^{iq_i\omega \delta t},
    \label{ASV}
\end{equation} 
here, $Nh$ represents the number of synthesized harmonics, and $q_i$ denotes the individual harmonic order. In Fig.~\ref{fig4}, we present a coherent superposition of harmonic vortex intensity distributions for different times corresponding to $\delta T=0.12 T$, $\delta = 0.25 T$, $\delta = 0.38 T$, and $\delta T = 0.5 T$, with $T$ being the driving-laser field period. We observe that the spatial distribution of the attosecond vortex beam rotates with time around the $\beta_x = \beta_x =0$ point. Unlike the case of a single harmonic, when a number of harmonic vortices are added coherently, the intensity distribution reflects an interference pattern that depends on the number of synthesized harmonics. Notice that, unlike other cases \cite{Garcia2024}, we are using scalar vortex beams rather than vectorial ones. Consequently, the resulting interference pattern arises from the addition of angular momentum of the different beams. By changing the number of synthesized harmonics, it is possible to shape the far-field spatial distribution of the attosecond vortex. The spectral phase for the time delays is shown in the inset of each panel. To plot the phase of the attosecond twisted pulse, we calculate the argument of $A_{as}$. This phase depends on the synthesized number of harmonics as well as the particular harmonics. This means that it will be possible to change the resulting attosecond harmonic vortex phase with the synthesized harmonic number. Notice also that the phase is time-dependent (see Eq.~(\ref{ASV})).

\begin{figure}[h!]
\includegraphics[width=0.8\textwidth]{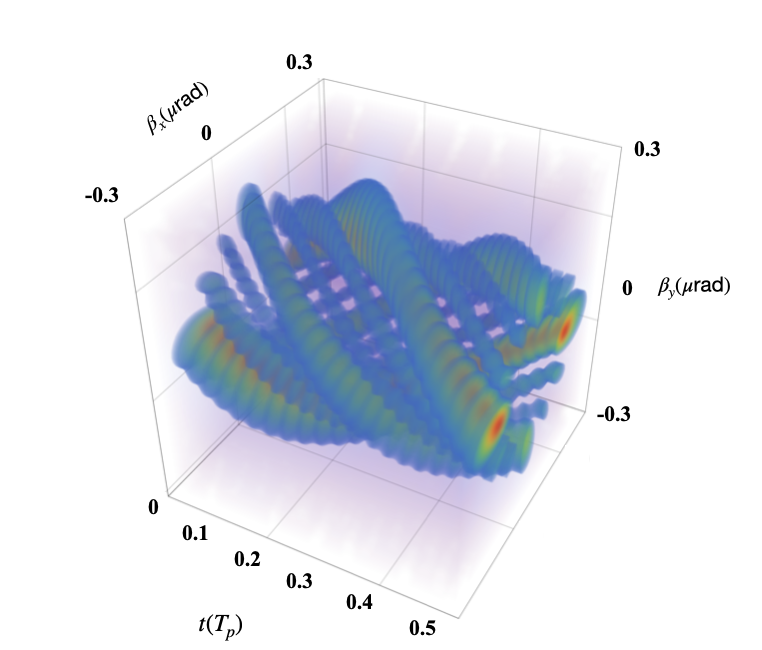}
\caption{Twisted attosecond pulse train. The plot shows the temporal evolution of the attosecond pulse formed by the superposition of different harmonic vortices. Here, $T_p$ represents the pulse period.}\label{fig5} 
\end{figure}

To visualize the spatial structure of the HHG emission, we plot the spatio-temporal evolution of a given HHG intensity. The simulated result is shown in Fig.\ref{fig5}. From this figure, we observe that a twisted attosecond pulse train is obtained, i.e., an attosecond pulse train delayed along the azimuthal coordinate $\phi$ according to the phase variation of the fundamental vortex beam. This spatio-temporal structure results from the superposition of harmonic vortices with different orbital angular momenta (OAMs).

It is well known that in order to generate attosecond pulse trains: (1) the spectral plateau in the HHG spectrum should closely resemble a frequency comb structure, and (2) the relative spectral phase between harmonics should remain nearly constant. As these conditions are met, along with the emission of vortex harmonics with almost identical divergence, the generation of a twisted attosecond pulse train is realized. Since the target material’s response and the fundamental beam structure are encoded in the twisted attosecond pulse train, the pulse train can be reshaped by tuning the OAM or wavelength of the fundamental beam.

From Fig.\ref{fig4}, it is clear that two main lobes rotate at different times within half a cycle of the laser period, accompanied by some side lobes. This feature can also be seen in Fig.\ref{fig5}. Thus, the number of intertwined helices in the twisted attosecond pulse train can be linked to the product of the difference between consecutive harmonic orders and the OAM of the fundamental beam. A similar feature has already been demonstrated for HHG driven by vortex beams carrying integer OAM in atomic gases \cite{Hernández:2015}. 

\section{Conclusions and Outlook}

In this work, we present a multi-scaling approach formed by the thin slab model (TSM) coupled with the Semiconductor Bloch Equations (SBE) to tackle the generation of high-order harmonics in solids. Our approach allows the successful combination of the microscopic aspects involved in harmonic generation in solids (described by the SBEs) with the macroscopic aspects resulting in the propagation of the generated radiation to the far field (described by the TSM). We are able to accurately reproduce experimental harmonic amplitude trends as a function of driving laser amplitudes. This is possible by appropriately accounting for the dephasing time, highlighting the significant role of correlations in harmonic spectra and their impact on attosecond vortex generation. Our findings elucidate the sensitivity of vortex beam divergence to amplitude scaling laws. Additionally, it is worth mentioning that by reproducing the experimental results in the plateau region of the HHG, we extend the validity of our theoretical approach to cover the entire harmonic spectra. We have previously demonstrated the validity of the approach in the nonlinear perturbative region. Due to the similar spatial divergence of the harmonic vortices and the generation of a long plateau (predicted by the SBE theory) that allows the synthesis of attosecond pulses, our theoretical exploration culminates in the demonstration of attosecond vortex beam generation. This represents a groundbreaking milestone as it is the first time such a field has been predicted in semiconductor materials. However, despite the simplicity of the model, the TSM plus SBE has been benchmarked along the harmonic spectra and presents a useful approach to describe the strong laser physics of monolayer materials. Other materials like MoS$_2$, graphene or topologically protected materials, which present different physics than bulk materials, are excellent candidates to extend the applicability of our model. The advent of attosecond twisted beams not only opens avenues for probing and inducing angular momentum coupling at the attosecond level in solid-state systems but also paves the way to carve XUV attosecond vortices.Finally, it is important to extend the work presented here to higher dimensions to explore the role of symmetries in the vortex beams.  Elucidating, enhancing and/or controlling material properties with optical vortices is an important quests that begins here by demonstrating that the vortex beams are sensitive to the dephasing time. 

\begin{acknowledgement}

We acknowledge financial support from the Guangdong Province Science and Technology Major Project (Future functional materials under extreme conditions - 2021B0301030005) and the Guangdong Natural Science Foundation (General Program project No. 2023A1515010871).
C.H. and S.G. acknowledge support by the U.S. Department of Energy, Office of Science, Basic Energy Sciences, Chemical Sciences, Geosciences, and Biosciences Division through the AMOS program.\end{acknowledgement}

\bibliography{achemso-demo}

\end{document}